\documentclass[prd,superscriptaddress,amsfonts,amssymb,onecolumn,amsmath,showpacs]{revtex4-2}
\usepackage{bm}
\usepackage{amsfonts}
\usepackage{latexsym}
\usepackage[utf8]{inputenc}
\usepackage{graphicx}
\usepackage{amsmath}
\usepackage{palatino}
\usepackage{mathpazo}
\usepackage[british]{babel}
\usepackage{hhline}
\usepackage{multirow}
\usepackage{textcomp}
\usepackage{float}
\usepackage{booktabs}
\usepackage{dcolumn}
\usepackage{lipsum} 
\usepackage{mdframed}
\usepackage[]{mdframed}
\usepackage{hhline}
\usepackage{multirow}
\usepackage{ragged2e}
\usepackage{hyperref}
\hypersetup{colorlinks,citecolor=blue}
\hypersetup{colorlinks=true,linkcolor=red,filecolor=magenta,    urlcolor=cyan}
\usepackage{amsmath}
\usepackage{xcolor}
\usepackage{orcidlink}
\usepackage{epsfig}
\usepackage{caption}
\usepackage{subcaption}
\usepackage{commath}

\def\jnl@style{\it}
\def\aaref@jnl#1{{\jnl@style#1}}

\def\aaref@jnl#1{{\jnl@style#1}}

\def\aj{\aaref@jnl{AJ}}                   
\def\apj{\aaref@jnl{ApJ}}                 
\def\apjl{\aaref@jnl{ApJ}}                
\def\apjs{\aaref@jnl{ApJS}}               
\def\apss{\aaref@jnl{Ap\&SS}}             
\def\aap{\aaref@jnl{A\&A}}                
\def\aapr{\aaref@jnl{A\&A~Rev.}}          
\def\aaps{\aaref@jnl{A\&AS}}              
\def\mnras{\aaref@jnl{Mon.~Not.~Roy.~Astron.~Soc.}}             
\def\prd{\aaref@jnl{Phys.~Rev.~D}}        
\def\prc{\aaref@jnl{Phys.~Rev.~C}}  
\def\prl{\aaref@jnl{Phys.~Rev.~Lett.}}    
\def\qjras{\aaref@jnl{QJRAS}}             
\def\skytel{\aaref@jnl{S\&T}}             
\def\ssr{\aaref@jnl{Space~Sci.~Rev.}}     
\def\zap{\aaref@jnl{ZAp}}                 
\def\nat{\aaref@jnl{Nature}}              
\def\aplett{\aaref@jnl{Astrophys.~Lett.}} 
\def\apspr{\aaref@jnl{Astrophys.~Space~Phys.~Res.}} 
\def\physrep{\aaref@jnl{Phys.~Rep.}}      
\def\physscr{\aaref@jnl{Phys.~Scr}}       
\def\commat{\aaref@jnl{Comm.~Math.~Phys.}}              
\def\science{\aaref@jnl{Science}}               
\def\cqg{\aaref@jnl{Classical Quant.~Grav.}}            
\def\jpcs{\aaref@jnl{JPCS}}                                     
\def\ijmpd{\aaref@jnl{Int.~J.~Mod.~Phys.~D}}                    
\def\grg{\aaref@jnl{Gen.~Relat.~Gravit.}}               
\def\rpp{\aaref@jnl{Rep.~Prog.~Phys.}}          
\def\npa{\aaref@jnl{Nucl.~Phys.~A}}        
\def\lrr{\aaref@jnl{Living Rev.~Rel.}}                   
\def\jcap{\aaref@jnl{J.~Cosmology Astropart.~Phys.}}    
\def\rmp{\aaref@jnl{Rev.~Mod.~Phys.}}   
\def\epjc{\aaref@jnl{Eur.~Phys.~J.~C}} 
\def\plb{\aaref@jnl{~Phy.~Lett.~B}} 
\def\mpla{\aaref@jnl{Mod.~Phy.~Lett.~A}} 
\def\arxiv{\aaref@jnl{arxiv.org}}

\allowdisplaybreaks[1]

\begin{document}

\color{black}       

\title{Cosmological Parameters in $f(T)$ Gravity: Theoretical and Observational Analysis}
\author{Suraj Kumar Behera\orcidlink{0009-0009-9294-460X}}
\email{skbehera.researches@gmail.com}
\affiliation{Department of Mathematics, School of Advanced Sciences, VIT-AP University, Beside AP Secretariat, Amaravati, 522241, Andhra Pradesh, India.}

\author{S. A. Kadam\orcidlink{0000-0002-2799-7870}}
\email{siddheshwar.kadam@dypiu.ac.in}
\email[]{k.siddheshwar47@gmail.com}
\affiliation{Centre for Interdisciplinary Studies and Research, D Y Patil International University, Akurdi, Pune-411044, Maharashtra, India}

\author{Pratik P. Ray \orcidlink{0000-0003-2304-0323}}
\email{pratik.chika9876@gmail.com}
\email[]{pratik.ray@vitap.ac.in}
\affiliation{Department of Mathematics, School of Advanced Sciences, VIT-AP University, Beside AP Secretariat, Amaravati, 522241, Andhra Pradesh, India.}
\affiliation{Pacif Institute of Cosmology and Selfology (PICS), Sagara, Sambalpur 768224, Odisha, India}

\author{B. Mishra\orcidlink{0000-0001-5527-3565}}
\email[]{bivu@hyderabad.bits-pilani.ac.in}
\affiliation{Department of Mathematics,
Birla Institute of Technology and Science-Pilani, Hyderabad Campus, Jawahar Nagar, Kapra Mandal, Medchal District, Telangana 500078, India.}

\date{\today}

\begin{abstract}
\textbf{Abstract}: The $f(T)$ gravity is one of the extensions of teleparallel equivalent of general relativity, in which more general functions of the torsion scalar $T$ can be described. With the proposed functional form of $f(T) = \alpha T - \beta u^{-n} + \gamma u^m$, where $u = (-T/6)$, we have analyzed the cosmological parameters using dynamical system analysis and cosmological datasets. The dynamical behavior of this model is analyzed with phase-space analysis by transforming the cosmological equations into an autonomous system. Critical points are identified, and their stability conditions examined, enabling the classifications of the early and late-time evolutionary phases of the Universe. The stability conditions are further demonstrated by phase-portrait diagrams that highlight transitions between radiation, matter, and dark-energy-dominated epochs. Then we used the Markov Chain Monte Carlo statistical technique to constrain the model parameters with the recent observational dataset, such as DESI DR2 BAO, and its combination with the Hubble and Pantheon+SH0ES data. The best-fit values for the model parameters were obtained by data analysis, $m \equiv 0.91^{+0.07}_{-0.09}$ and $n \equiv 0.69^{+0.09}_{-0.08}$, and are well within the stability range obtained ($m<1\land n>-1$) through dynamical system analysis. The combined theoretical and observational analysis shows that the proposed $f(T)$ gravity model successfully reproduces the observed cosmic expansion history of the Universe.
\end{abstract}

\maketitle
\textbf{Keywords}: $f(T)$  gravity, critical points, cosmological datasets, evolutionary behavior.

\section{Introduction}\label{Introduction}
Cosmological investigations, including large-scale structure \cite{Tegmark_2004, Eisenstein_2005}, cosmic microwave background radiation (CMBR)\cite{Spergel_2003,2007ApJS..170..377S}, Type Ia supernovae (SNeIa)\cite{Riess_1998, Perlmutter_1999} have revealed the accelerating behavior of the Universe. The majority think that the principal gradient accelerating the cosmos is a mysterious component known as dark energy (DE). The Universe is composed of cold dark matter (CDM) and is governed by this strange energy component, defined by negative pressure. It also propels the current acceleration of the expansion of the Universe. In the literature, several DE models have been proposed, such as holographic DE \cite{Cohen_1999, LI20041}, new agegraphic DE\cite{WEI20081, WEI2008113}, Ricci DE\cite{Gao_2009}, quintessence, phantom, k-essence tachyon, and quintom\cite{Caldwell_1998, Caldwell_2002, Armendariz_Picon_2001, Padmanabhan_2002, AshokeSen_2005, Elizalde_2004, 
FENG200535}. In addition, the Chaplygin gas\cite{KAMENSHCHIK2001265}, the generalized Chaplygin gas (GCG)\cite{Bento_2002}, and so forth. One of the most difficult problems in modern cosmology and particle physics is the nature of DE. It is estimated that DE constitutes about 75\% of the total energy of the Universe. The simplest and most attractive choices for DE are the vacuum energy (cosmological constant $\Lambda$) and the constant equation-of-state (EoS) parameter $\omega\approx-1$. The minuscule observed value of DE density resulting from quantum field theories becomes difficult to explain, which is called the cosmological constant problem \cite{Carroll_2001}. 

There are two possibilities to explain the accelerating behavior of the Universe. The first one involves adding new fields, such as vector fields, phantom scalars, canonical scalars, and so on \cite{copeland, Bassett_2006, CAI20101}, to alter the matter component of the Universe. Whereas, modifying the gravitational sector is the second possibility \cite{Capozziello_2011, Di_Valentino_2025}. The Einstein-Hilbert action, which is based on curvature, is usually broadened to formulate modified gravitational theories; however, an alternative class of gravitational modification can be obtained by broadening the action of the equivalent torsion formulation of general relativity, known as the teleparallel equivalent of GR (TEGR) \cite{golovnev2018introductionteleparallelgravities}.  In TEGR, torsion is used as the gravitational interaction and tetrad fields serve as the fundamental dynamical quantities. This provides an orthonormal set of four linearly independent vector bases in the tangent space associated with each spacetime. The modification of TEGR, referred to as $f(T)$ gravity, has emerged as a promising framework to describe late time cosmic phenomena of the Universe. In the formulation of $f(T)$ gravity \cite{Ferraro_2008, Bengochea_2009, Linder_2010}, pure tetrad teleparallel gravity is the starting point, and the spin connection is taken to vanish. Consequently, the torsion tensor is described by connection coefficients linked to the governing parameters, and these quantities fail to obey local Lorentz transformations. Since the field equations remain unaffected by the breaking of local Lorentz symmetry, it has been ignored in TEGR. The teleparallel gravity (TG) makes use of the teleparallel connection $\Gamma^\rho_{\mu \nu}$ \cite{Krssak:2018ywd, Bahamonde_2023}, which has vanishing curvature but having torsion. 

Some of the results of cosmology and astrophysics problems in $f(T)$ gravity are available in the literature such as the late time phenomena  \cite{Ferraro_2008, Bengochea_2009, Linder_2010, 2011EPJC...71.1752M,James_B._Dent_2011}, models with observational constraints  \cite{BENGOCHEA2011405, WU2010415, zhang2011notes, WEI201174}, the large-scale structure \cite{WEI201174} and so on. The cosmological perturbation has been analyzed in Ref. \cite{Chen_2011, Cai_2011, shivam_2025, Shivam_2025_2}. The finite-time future singularity in $f(T)$ gravity has been shown in \cite{reconsturction}, whereas the  Noether symmetry approach to find an exact solution to the given Lagrangian is significant  \cite{WEI2012298,Jamil_2012}. Zheng and Huang  \cite{Rui_Zheng_2011} have compared the power-law framed model and the observable prediction of $\Lambda$CDM and DE models. Setare and Mohammedipour \cite{M.R._Setare_2013} have shown the cosmological dynamics that comply with $\Lambda$CDM cosmic history. The spherically symmetric static solutions in $f(T)$ theories investigated in Ref. \cite {PhysRevD.84.024042} and the cosmic expansion with cosmographic parameters has been shown \cite{Capozziello_2011}. The phase-space dynamical analysis of the teleparallel DE concept has been conducted using TEGR \cite{Chen_Xu_2012}. The phantom divide cannot be crossed in the logarithmic type $f(T)$ model, but it can be realized in the combined logarithmic and exponential terms\cite{Kazuharu_Bamba_2011}. Two $f(T)$ models that realize the phantom-divide crossings for the EoS parameter have also been proposed by Wu and Yu \cite{2011EPJC...71.1552W}. Data sources, including the Pantheon supernova sample, Hubble constant measurements, the cosmic microwave background shift parameter, and redshift-space distortion measurements, have been employed to constrain $f(T)$ gravity in \cite{Anagnostopoulos_2019}. Ref.\cite{Said_2020} addresses the effects of the electromagnetic sector violation of the equivalence principle on $f(T)$ gravity. Ref.\cite{Cai_2020, Briffa_2020} shows how the $f(T)$ Lagrangian was reconstructed using Hubble data and background cosmological parameters. The same methodology was then applied to the growth rate data to constrain the $f(T)$ Lagrangian values \cite{Levi_Said_2021}. The dynamical stability methodology we have adopted in the present work for analyzing autonomous dynamical systems in the $f(T)$ gravity formalism has been discussed in Refs.\cite{zhang2011notes, Mirza_2017,Duchaniya:2022rqu,Duchaniya_2024}.

With dynamical system analysis, the cosmic dynamics and asymptotic behavior of the cosmological models can be analyzed successfully. One can avoid the nonlinearities involved in the cosmological equations \cite{zhang2011notes}. The stability of the cosmological model can be addressed through the eigenvalues and phase portrait. From a theoretical perspective, Inflation $\rightarrow$ Radiation $\rightarrow$ Matter $\rightarrow$ DE era should be explained by at least some of the suggested cosmological models \cite{bohmer2017dynamicalsystem,Kadamdynamicalftb}. To achieve this proposed cosmological model, unstable radiation and matter points must be saddle for these periods to be sufficiently long \cite{Bahamonde_2023}. At the same time, inflation must be an unstable point for the Universe to have an inflation exit. The DE era is at the latter stage, a steady time of rapid growth, and can be described through a stable attractor. Additionally, a description of global dynamics can be achieved by linking critical points to relevant evolutionary epochs \cite{zhang2011notes, Mirza_2017,duchaniya2023dynamical, Kadam:2022lgq}. With this motivation, we study the dynamical stability $f(T)$ gravity model with some well defined form of the functional $f(T)$. The paper is organized as follows: in Sec. \ref{sec:cwfrl}, the field equation for $f(T)$ gravity is derived. In Sec. \ref{dynamical_section}, the dynamical stability of the model is discussed. In Sec. \ref{data_sec}, the comparison of the model with observational data has been discussed. The concluding remarks are presented in Sec. \ref{conclusion}.

\section{Basic formalism of $f(T)$ gravity}
\label{sec:cwfrl}
In this section, we present the mathematical formalism of $f(T)$ gravity. The vierbein fields, denoted as $e_i(x^{\mu})$, operate as non-holonomic frames in the context of spacetime and constitute the primary dynamical variables in teleparallel gravity. At each point $x^{\mu}$ on the manifold, the vierbein fields provide an orthonormal set of basis vectors in the tangent space, ensuring that $g(e_i, e_j) = \eta_{ij}$, with $\eta_{ij}$ representing the Minkowski metric of flat spacetime. In a coordinate framework, the vierbeins can be expressed as $e_i = h^{\mu}_{i} (x)\partial_{i}$, leading to the definition of the metric tensor \cite{Capozziello_2011, Paliathanasis:2016vsw},

\begin{equation}
g_{\mu\nu}(x) = \eta_{ij} h^{i}_{\mu}(x) h^{j}_{\nu}(x)\,.
\end{equation}
In teleparallel gravity, instead of the curvature that defines the Levi-Civita connection, the torsion is used for the Weitzenböck connection and can be expressed as \cite{Hayashi:1979qx},
 
\begin{equation}
T^{\lambda}_{\mu\nu} = \Gamma^{\lambda}_{\nu\mu} - \Gamma^{\lambda}_{\mu\nu} = h^{\lambda}_{i} ( \partial_{\mu} h^{i}_{\nu} - \partial_{\nu} h^{i}_{\mu} )\,.
\end{equation}
The gravitational field is described by a Lagrangian density that depends on the torsion scalar $T$. Consequently, the action is expressed as \cite{Anagnostopoulos_2019}, 
\begin{equation} \label{action}
    S=\frac{1}{16 \pi G}\int d^4xe\left[T+f\left(T\right)+\mathcal{L}_m\right],
\end{equation}
where, $e=det(e^{i}_{\mu})$ and $\mathcal{L}_m$ is the matter Lagrangian. The contraction of the superpotential and torsion tensors results in the torsion scalar $T$ and can be expressed as,

\begin{equation}\label{torsion_scalar}
T = S_{\rho}^{\; \mu\nu} T^{\rho}_{\; \mu\nu}\,,
\end{equation}
in which the superpotential takes the form,

\begin{equation}
S_{\rho}^{\; \mu\nu}
= \frac{1}{2} \left(
K^{\mu\nu}_{\;\;\;\;\rho}
+ \delta^{\mu}_{\rho} T^{\theta\nu}_{\;\;\;\;\theta}
- \delta^{\nu}_{\rho} T^{\theta\mu}_{\;\;\;\;\theta}
\right)\,.
\end{equation}
The contortion tensor becomes,

\begin{equation}
K^{\mu\nu}_{\;\;\;\;\rho}
= -\frac{1}{2} \left(
T^{\mu\nu}_{\;\;\;\;\rho}
- T^{\nu\mu}_{\;\;\;\;\rho}
- T_{\rho}^{\; \mu\nu}
\right)\,.
\end{equation}
Varying the action [Eq. \eqref{action}] with respect to the tetrad provides the field equation as \cite{Bengochea_2009,Bahamonde_2023},

\begin{equation}\label{general_field_equation}
e^{-1}\partial_{\mu}\!\left( e\, e^{\rho}_i S^{\;\;\mu \nu}_{\rho} \right)\!\left[ 1 + f_T \right]
+ e^{\rho}_i S^{\;\;\mu \nu}_{\rho}\,\partial_{\mu}(T) f_{TT} \\
- e^{\lambda}_i T^{\rho}_{\;\;\mu \lambda} S^{\;\;\nu \mu}_{\rho}\!\left[ 1 + f_T \right]
+ \frac{1}{4} e^{\nu}_i \big[ T + f(T) \big]
= 4\pi G\, e^{\rho}_a T^{\;\;\nu}_{\rho}.
\end{equation}
Here, $f_T$ and $f_{TT}$ respectively represent the first and second derivatives of the function $f(T)$ with respect to the torsion scalar $T$. The tensor $T^{~~\mu}_{\rho}$ corresponds to the energy-momentum tensor associated with the matter source $S_m$. Note that,Eq. \eqref{general_field_equation} reduces to GR for the choice $f(T) = 0$ substituted into the action given in Eq. \eqref{action}. To frame the cosmological model, we consider the background as flat, homogeneous, and isotropic, and can be represented by the tetrad as \cite{Krssak:2018ywd},

\begin{equation}\label{flrw_tetrad}
    e^{a}_{\ \ \mu}=\textrm{diag}(1,a(t),a(t),a(t))\,,
\end{equation}
where $a(t)$ represents the scale factor as a function of cosmic time $t$. It is to note that the standard flat Friedmann-Lemaître-Robertson-Walker (FLRW) metric can be obtained by using the connection between the metric and the tetrad. So, the line element expressed in Cartesian coordinates as \cite{misner1973gravitation},

\begin{equation}\label{st}
ds^2 = -dt^2 + a^2(t) (dx^2 + dy^2 + dz^2)\,.
\end{equation}
For the unperturbed FLRW metric \eqref{st}, the torsion scalar \eqref{torsion_scalar} becomes

\begin{equation}
T = -6 \left( \frac{\dot{a}}{a} \right)^2 = -6H^2\,,
\end{equation}
and hence the field equations of $f(T)$ gravity obtained as

\begin{align}
3H^2 &= 8 \pi G \rho -\frac{f(T)}{2}+T f_T \,,\label{1st_Friedmann}\\
\dot{H} &= -\frac{4 \pi G \left(\rho + p \right)}{1 + f_T + 2 T f_{TT}} \,. \label{2nd_Friedmann}
\end{align}
From Eq. \eqref{1st_Friedmann} and Eq. \eqref{2nd_Friedmann}, one can separate out the contribution of DE components of energy density and pressure,

\begin{align}
\rho_{DE} &= \frac{1}{16 \pi G} \left[-f + 2 T f_T\right] \,,\\
 p_{DE} &= -\frac{1}{16 \pi G} \left[\frac{-f + T f_T - 2 T^2 f_{TT}}{1 + f_T + 2 T f_{TT}}\right] \,,
\end{align}
and subsequently the EoS parameter for DE becomes,

\begin{align}\label{eosde}
 \omega_{DE} = -1 + \frac{(f_T + 2 T f_{TT}) (-f + T + 2 T f_T)}{(1 + f_T + 2 T f_{TT}) (-f + 2 T f_T)}\,.
\end{align}
This set of field equations [Eq.\eqref{1st_Friedmann}--Eq.\eqref{2nd_Friedmann}] will be examined with the following well-motivated form of the function $f(T)$ \cite{Yang:2011},
\begin{equation}\label{ftform}
f(T) =\alpha T - \beta u^{-n} + \gamma u^m,
\end{equation}
where $\alpha$, $\beta$, $\gamma$, $n$ and $m$ are model parameters and $u = (-T/6)$. In the following sections, we shall perform the dynamical stability of the cosmological model with phase space analysis and the analysis of cosmological parameters using observational data sets such as $H(z)$, DESI DR2 BAO, and Pantheon+SH0ES.

\section{The Phase Space Analysis}\label{dynamical_section}
For the qualitative analysis of cosmological models, dynamical system analysis has been an effective tool \cite{Franco:2021, Franco:2020lxx}. The analysis is important since modified gravity theories result in complex systems of differential equations characterized by ambiguous initial conditions. This complicates the search for analytical solutions for the system. This method enables us to examine the long-term dynamics of the system and helps in identifying the regions of instability and attractors. The cosmological systems can represent various evolutionary phases, and the asymptotic behavior at late times tends to converge \cite{zhang2011notes, Mirza_2017}. This behavior can be established through a stable critical point derived from an autonomous system generated from the cosmological equations, with earlier phases characterized by unstable nodes or saddle points. Here, we shall perform the analysis with the form $f(T)$ as in Eq. \eqref{ftform}. The partial differentiation with respect to the torsion scalar $T$ can be obtained as, 

\begin{align}
f_T &= \alpha - \frac{\beta n}{6}u^{-(n+1)} - \frac{\gamma m}{6}u^{m-1} \,,\\
f_{TT} &= -\frac{\beta n(n+1)}{36}u^{-(n+2)}
+
\frac{\gamma m(m-1)}{36}u^{m-2}\,.
\end{align}
Now, the first Friedmann equation \eqref{1st_Friedmann} becomes,
\begin{equation}
\frac{\kappa^2 \rho_m}{3H^2} + \frac{\kappa^2 \rho_r}{3H^2} + \frac{\beta u^{-n}}{3H^2}
\left(\frac{1}{2}+n\right) -
\left[
\frac{\gamma u^{m}}{3H^2}
\left(\frac{1}{2}-m\right)
+\alpha
\right]= 1\,.
\end{equation}
We substitute,
\begin{align}
\Omega_m = \frac{\kappa^2 \rho_m}{3H^2},~~~ \quad r = \frac{\kappa^2 \rho_r}{3H^2},~~~ \quad  x =
\frac{\beta u^{-n}}{3H^2}
\left(\frac{1}{2}+n\right),~~~ \quad y =
-\left[
\frac{\gamma u^{m}}{3H^2}
\left(\frac{1}{2}-m\right)
+\alpha
\right]\,,
\end{align}
are dimensionless variables that will be used to form the autonomous dynamical system. These variables satisfy the following constraint equation,

\begin{equation}
\Omega_m + r + x + y = 1\,. \label{constrain_eq}
\end{equation}
The autonomous dynamical system can be obtained by differentiating the dimensionless variables with respect to the e-folding number $N=ln(a)$ and the system becomes,

\begin{align}
x' &= -2(n+1)x \frac{\dot{H}}{H^2} \,,\notag\\
y' &= 2(m-1)(\alpha+y) \frac{\dot{H}}{H^2} \,,\notag\\
\gamma' &= -2r\left(2 + \frac{\dot{H}}{H^2}\right) \,.\label{dynamical_system}
\end{align}
Using the second Friedmann equation \eqref{2nd_Friedmann}, we can obtain,

\begin{equation}
\frac{\dot{H}}{H^2} = \frac{\frac{3}{2}(x+y-1) - \frac{r}{2}}{1 + nx - my+\alpha(1-m)}\,.
\end{equation}
The EoS parameters and deceleration parameter can be expressed in the form of the dimensionless variables as,

\begin{align}
\omega_{DE} &=-1 +
\frac{(n x + m y + \alpha (1-m))(x + y - 1)}
{(n x - m y + 1 + \alpha (1-m))(x
+ y)}\,,\\
\omega_{tot} &= -1 -
\frac{(x + y - 1) -\frac{r}{3}}
{1 + n x - m y + \alpha (1-m)} \,,\\
q &= -1+\frac{r-3 (x+y-1)}{2 (\alpha -m (\alpha +y)+n x+1)}\,.
\end{align}
To obtain the dynamical features of the autonomous system, we need to solve the coupled equations $x'=0$, $y'=0$ and $r'=0$. The corresponding critical points and their eigenvalues of the system \eqref{dynamical_system} are given in Table--\ref{table_critical_pont}. The stability of the critical point is characterized by the signs of the eigenvalues of the Jacobian matrix at the critical point. For more details, one may see Refs. \cite{Gonzalez-Espinoza:2020jss, Kadam:2024fTBTGBG, Kadam:2024Tele}.

\begin{table*}[!htb]
\small
\addtolength{\tabcolsep}{-2pt}
\begin{tabular}{|c|c|c|c|c|}
\hline
\textbf{Critical Points} 
& \textbf{$x$} 
& \textbf{$y$} 
& \textbf{$r$} 
& \textbf{Eigenvalues} \\
\hline
\hline
$A_1$ 
& $x$ 
& $1-x$ 
& $0$ 
& $\{0,-4,-3\}$ \\
\hline
$A_2$ 
& $0$ 
& $-\alpha$ 
& $1+\alpha$ 
& $\{1,-4 (m-1),4 (n+1)\}$ \\
\hline
$A_3$ 
& $0$ 
& $-\alpha$ 
& $0$ 
& $\{-1,3-3 m,3 (n+1)\}$\\
\hline
\end{tabular}
\caption{Critical points for the dynamical system  [$(x,y,r)$ phase space]}
\label{table_critical_pont}
\end{table*}

\begin{figure}[ht]
    \centering
    \includegraphics[width=0.35\textwidth]{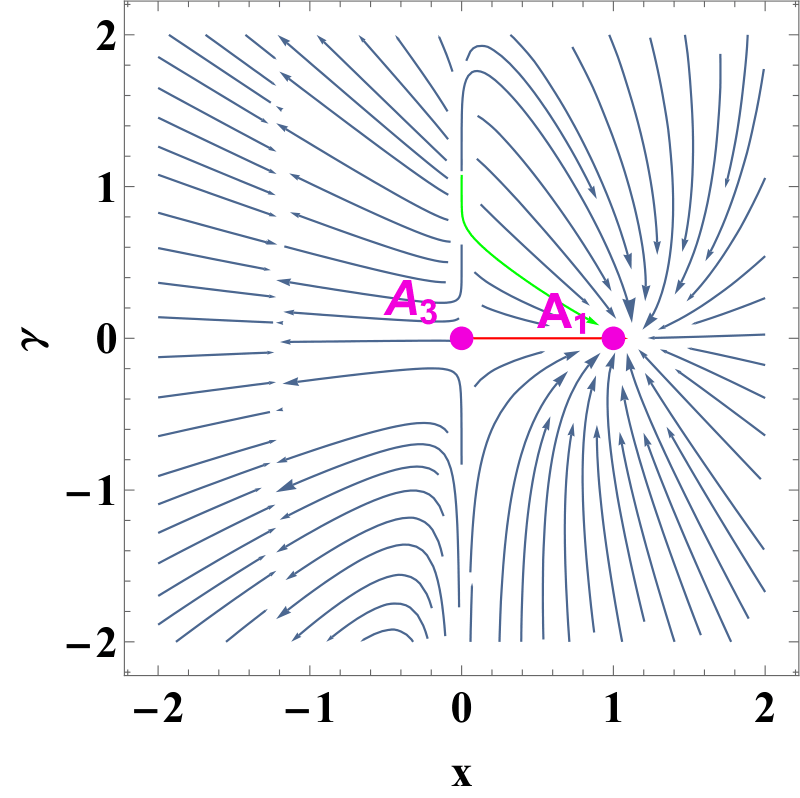}
    \includegraphics[width=0.37\textwidth]{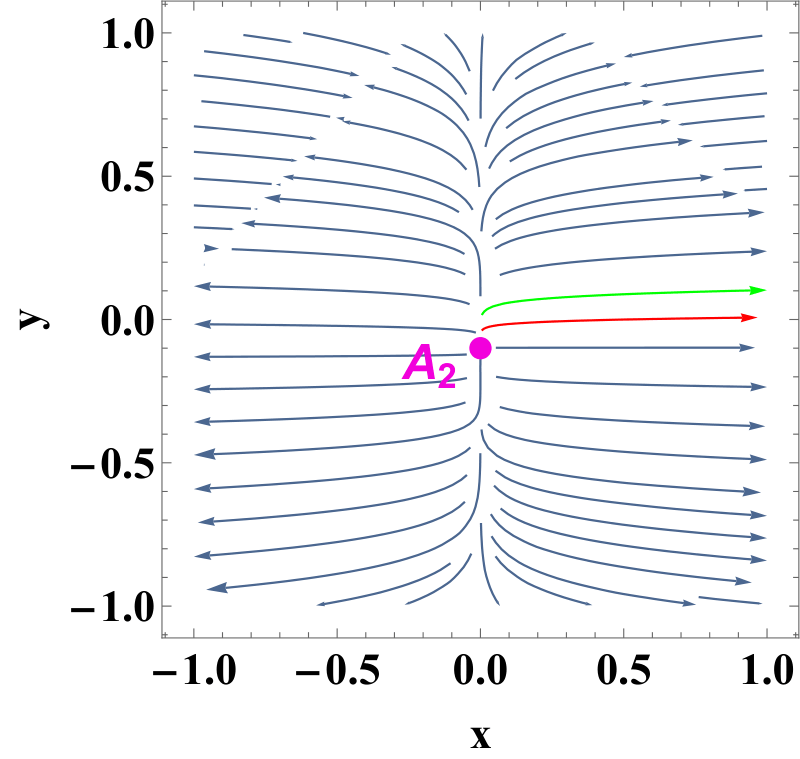}
    \caption{
    Phase space diagram of the autonomous system showing the critical points
    $A_1$, $A_2$, and $A_3$.
    }
    \label{fig:phase_space_A}
\end{figure}

\begin{itemize}
    \item{\bf Critical Point $A_1$:} This is a critical point representing the DE-dominated epoch with $q= \omega_{tot} =-1$ and is a de-Sitter solution [Table--\ref{stability_table}]. The eigenvalues at this critical point is non-hyperbolic with one vanishing eigenvalues. The coordinates of this critical point show that it forms a one-dimensional non-isolated set, appropriate with the presence of a single zero eigenvalues.  Hence, this critical point is hyperbolic \cite{Khyllep:2023fQ}. The stability depends on the sign of the remaining eigenvalues; here, both the remaining eigenvalues are negative, hence this critical point is stable. As depicted in Fig.--\ref{fig:phase_space_A}, the phase-space trajectories near this critical point exhibit convergence, indicating its function as a late-time attractor. Here, the values of density parameters at this critical point are $\Omega_{r}=0$, $\Omega_{DE}=1$, $\Omega_{r}=0$ reflecting that the critical point is DE-dominated.
    
    \item{\bf Critical Point $A_2$:} At this point, the Universe evolves in a radiation regime, indicated by $q=1$ and $\omega_{tot} = \frac{1}{3}$. The stability criteria are listed in Table--\ref{stability_table}, while the nature of the critical point is derived from the eigenvalue signature summarized in Table--\ref{table_critical_pont}. Fig.--\ref{fig:phase_space_A} demonstrates the corresponding phase-space behavior in the $x-y$ plane. This shows that the behavior of phase-space trajectories at this critical point is moving away from it, indicating the instability of this critical point. The same is encountered from the eigenvalues, which shows instability for $m<1\land n>-1$. This critical point is a non-standard radiation-dominated critical point with $\omega_{r}=1+\alpha$. There is a small contribution from the DE at this critical point with $\Omega_{de}=-\alpha$ and no contribution from the matter side $\Omega_{m}=0$. 
    
 \item{\bf Critical Point $A_3$:} This is a critical point representing a matter-dominated epoch with $q=\frac{1}{2}, \omega_{tot}=0$. Stability properties of the critical point is defined by examining the eigenvalues signs listed in Table--\ref{stability_table}. This critical point shows saddle point behavior for the model parameter range $m<1\land n>-1$, $m>1\land n>-1$, $m<1\land n<-1$. We obtained the model parameter range $m>1\land n>-1$ to plot the phase space diagrams, which has been presented in Fig.--\ref{fig:phase_space_A}. The trajectories are moving away from this critical point, showing the saddle point behavior. Along with the saddle point behavior, this matter-dominated critical point shows stable behavior within the range $m>1\land n<-1$. This critical point will behave as a standard matter-dominated critical point at $\alpha=0$. The values of density parameters at this critical point are $\Omega_{m}=1+\alpha, \Omega_{r}=-\alpha$ and $\Omega_{de}=0$.
\end{itemize}

\begin{table*}[!htb]
\centering
\small
\addtolength{\tabcolsep}{-2pt}
\begin{tabular}{|c|c|c|c|c|c|c|}
\hline
\textbf{C.P.} 
& \textbf{Stability} 
& \textbf{$q$} 
& \boldmath{$\omega_{\text{tot}}$}
& \boldmath{$\Omega_m$} 
& \boldmath{$\Omega_r$} 
& \boldmath{$\Omega_{de}$} \\
\hline
\hline
$A_1$ 
& Stable
& $-1$ 
& $-1$
& $0$ 
& $0$ 
& $1$ \\
\hline
$A_2$ 
& \begin{tabular}{@{}c@{}} 
Unstable for \\ 
$m<1\land n>-1$,\\ 
Saddle for \\ 
$m>1\land n<-1,$ \\ 
$m<1\land n<-1,$ \\
$m>1\land n>-1\,.$
\end{tabular}
& $1$ 
& $\dfrac{1}{3}$ 
& $0$ 
& $1+\alpha$ 
& $-\alpha$ \\
\hline
$A_3$ 
& 
\begin{tabular}{@{}c@{}}
Saddle for\\
$m<1\land n>-1$, \\
$m>1\land n>-1$,\\
$m<1\land n<-1$.
\end{tabular}
& $\dfrac{1}{2}$ 
& $0$ 
& $1+\alpha$ 
& $-\alpha$
& $0$\\
\hline
\end{tabular}
\caption{The stability conditions, deceleration parameter, EoS parameters, and density parameters.}
\label{stability_table}
\end{table*}

\begin{figure}[htbp]
    \centering
    \begin{minipage}{0.32\textwidth}
        \centering
        \includegraphics[width=\textwidth]{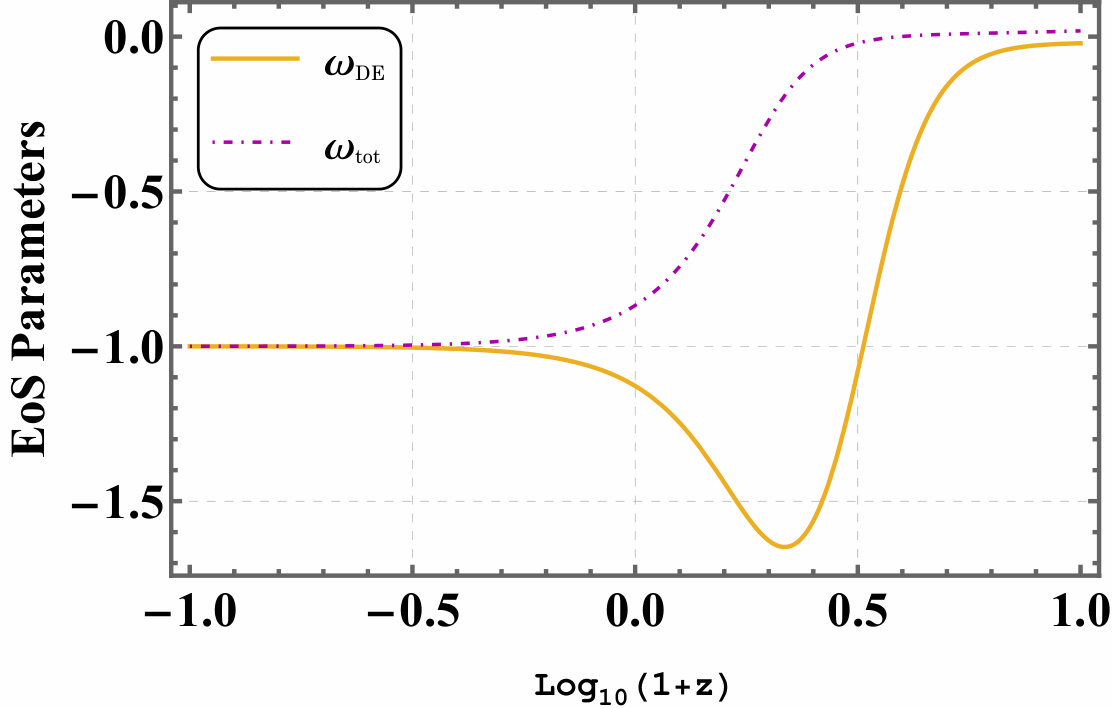}
        \vspace{2pt}

    \end{minipage}\hfill
    \begin{minipage}{0.32\textwidth}
        \centering
        \includegraphics[width=\textwidth]{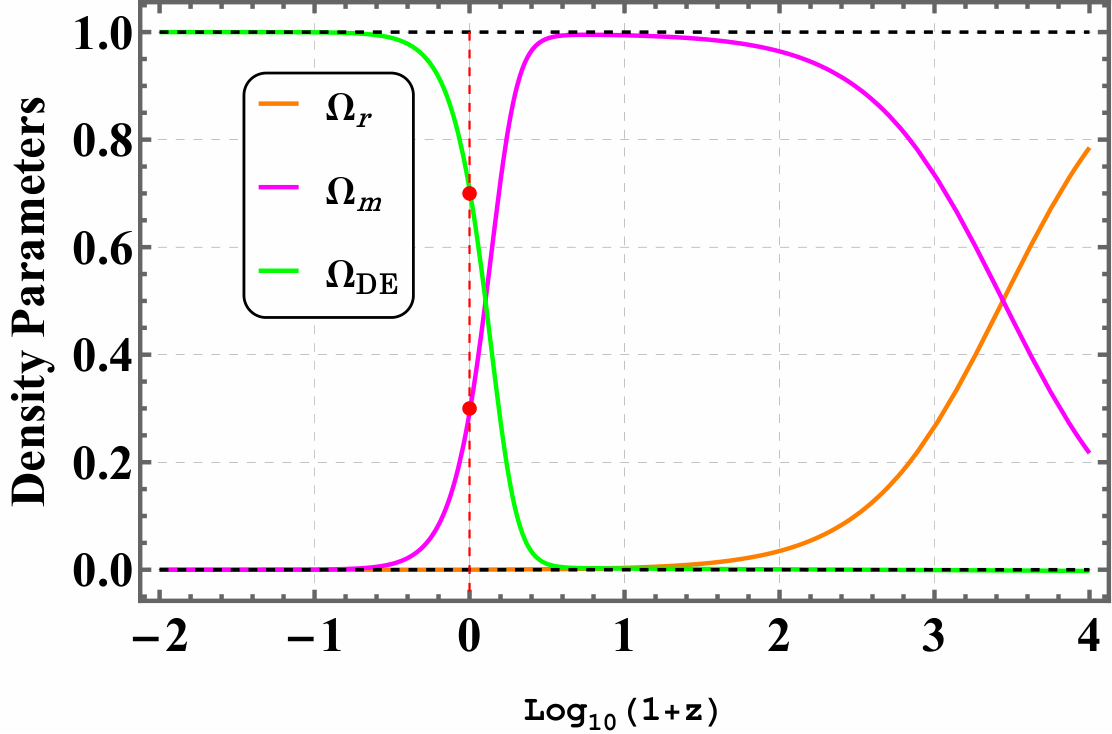}
        \vspace{2pt}
    \end{minipage}\hfill
    \begin{minipage}{0.32\textwidth}
        \centering
        \includegraphics[width=\textwidth]{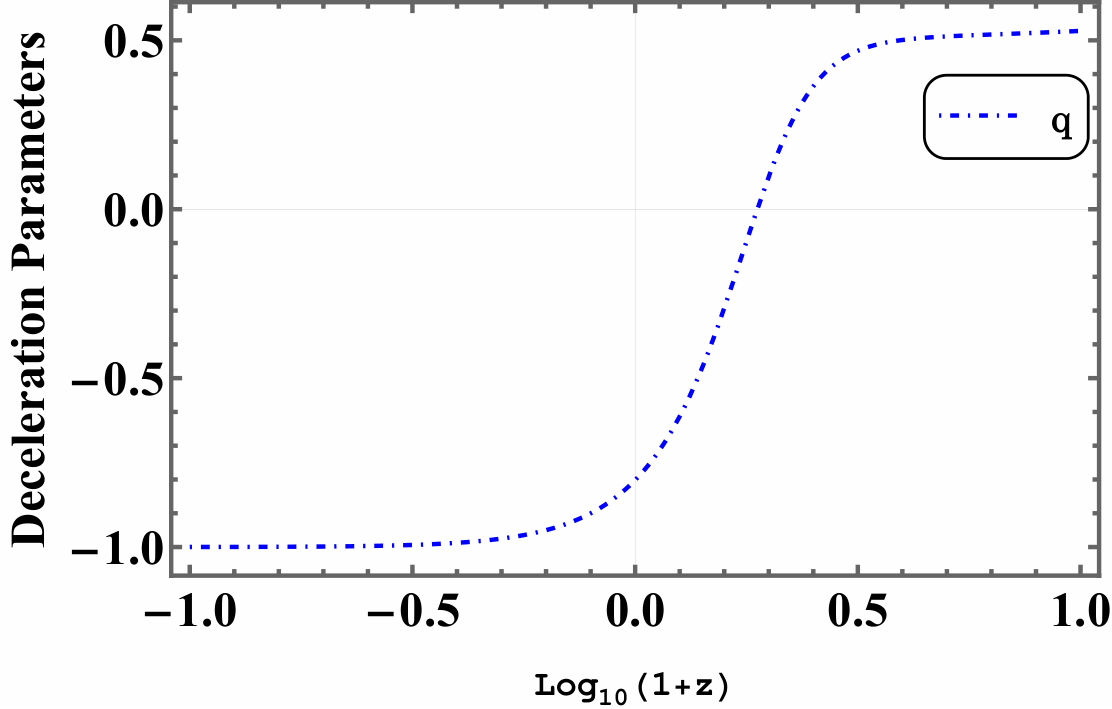}
        \vspace{2pt}
    \end{minipage}
    \caption{The evolution of EoS parameters, density and deceleration parameter as functions of redshift with the initial condition $x_0=10^{-12}, y_0 = 0.674, r_0=9.1 \times 10^{-23}$.}
    \label{fig_evolution_plot}
\end{figure}
In Fig.--\ref{fig_evolution_plot}, we present a comparison of EoS parameters pertaining to DE ($\omega_{DE}$) and total energy density ($\omega_{tot}$). The evolutionary behavior of both the EoSs are approaching to $-1$ in the late cosmological epoch. The value of $\omega_{DE}$ at $z=0$ is $-1.015$, which slightly crosses $-1$ and is in the phantom region, aligns with the observational findings from the Planck Collaboration \cite{Aghanim:2018eyx}. We illustrate the changes in energy densities for radiation, DE, and DM through the behavior of density parameters. These plots reveal that radiation was the dominant component during the early evolutionary phases. Furthermore, it emphasizes the short period where the Universe was predominantly matter-dominated, followed by the rise of the cosmological constant in the late Universe. The present value of the matter density ($\Omega_{m}$) and DE density ($\Omega_{DE}$) obtained respectively, $\Omega_{m}$ $\approx 0.3$ and $\Omega_{DE} \approx 0.7$. The time of matter-radiation equality, marked by a pointed arrow in Fig.--\ref{fig_evolution_plot}, occurs at approximately $z \approx 3387$. In Fig.--\ref{fig_evolution_plot}, we investigate the characteristics of the deceleration parameter. The graph indicates a transient behavior at $z \approx 0.66$, which aligns with the findings from \cite{PhysRevD.90.044016a} and the present value of the deceleration parameter $q \approx -0.53$ \cite{PhysRevResearch.2.013028}.

\section{The Observational Analysis}\label{data_sec}
From the dynamical system analysis, we precisely obtained the range of the model parameters. To further validate this result, we shall confront this with the cosmological datasets. Substituting the form of $f(T)$ in Eq. \eqref{1st_Friedmann} and incorporating the initial condition of the present time, we obtain

\begin{equation} \label{alpha}
    \beta = \frac{H_0^{2n}\left[ 6H_0^2 \left( 1-\Omega_{m0}-\Omega_{r0} \right)\alpha 6 H_0^2-H_0^{2m} (2m-1) \gamma \right]}{2n+1},
\end{equation}
where $H_0$, $\Omega_{m0}$, and $\Omega_{r0}$ respectively denote the Hubble parameter, matter density, and radiation density at the present time. From Eq. \eqref{alpha} and Eq. \eqref{1st_Friedmann}, an implicit form of the Hubble parameter can be obtained as,

\begin{equation}
\begin{aligned}
H^2 &= \frac{1}{6} H^{-2 n} \Big( \gamma (2 m-1) H^{2 m+2 n}
-6 \alpha H^{2 n+2}  \\
&\quad + H_0^{2 n} \big(6 \alpha H_0^2
- \gamma (2 m-1) H_0^{2 m}
+ 6 H_0^2 (1-\Omega_{\text{m0}}-\Omega_{\text{r0}}) \big) \Big)  \\
&\quad + H_0^2 \left((1+z)^3 \Omega_{\text{m0}}+(1+z)^4 \Omega_{\text{r0}}\right)
\end{aligned}\label{Hubble}
\end{equation}

Being an implicit equation, the analytical solution for Eq. \eqref{Hubble} is unfeasible; hence, we resort to the numerical approach to find the constrained values of the parameters, $H_0$, $\alpha$, $\gamma$, $\Omega_{m0}$, $m$, and $n$. 

To begin with the observational analysis, we employ the {\it emcee} package to execute Markov Chain Monte Carlo (MCMC) analysis, integrating various cosmological observational datasets. The MCMC is an effective statistical sampling approach used to provide samples from the posterior probability distribution of cosmological model parameters. This method is used in the Bayesian statistical framework to derive robust bounds on the model parameters and their associated uncertainties. We shall use observational datasets such as Hubble parameter $H(z)$ measurements, Pantheon+SH0ES Type Ia supernova compilation, and the DESI DR2 Baryon Acoustic Oscillation (BAO) data. A brief discussion on each of the cosmological datasets to be used is presented below. 

The observational  $H(z)$ datasets offer significant evidence for understanding the cosmic expansion history. The entire list of the datasets is presented in \cite{BHAGAT2023101250}. In this study, we constrain the model parameters by a $\chi^2$ statistical analysis performed within the MCMC framework. The $\chi^2$ function associated with the $H(z)$ dataset is defined as, 

\begin{equation}
\chi^{2}_{\mathrm{\textit{H(z)}}} = \sum_{i=1}^{32} \frac{\left[ H_{\mathrm{th}}(z_i) - H_{\mathrm{obs}}(z_i) \right]^2}{\sigma_i^2}\,,
\end{equation}
where $\sigma_i$ is the standard error associated with the observational measurement of the Hubble parameter. The quantities $H_{\text{th}}(z_i)$ and $H_{\text{obs}}(z_i)$ respectively denotes the theoretical prediction and the observed value of the Hubble parameter at redshift $z_i$ .

Type Ia Supernova (SNIa) datasets have been significant in the investigation of late-time cosmic expansion. Owing to their nearly uniform intrinsic luminosity, they function as reliable standard candles, enabling high-precision luminosity-distance evaluation across extensive redshift coverage. The {\bf Pantheon+SH0ES dataset} contains 1701 light-curve measurements corresponding to 1550 supernovae in the redshift range $0.00122 < z < 2.2613$ \cite{Brout_2022}. For a given cosmological model, the theoretical distance modulus is defined as, 

\begin{equation}
\mu_{\mathrm{th}}(z,\theta) = 5 \log_{10}\left(d_L(z,\theta)\right) + 25,
\end{equation}
where $\theta = [H_0, \alpha, \gamma, \Omega_{m0}, m, n]$ denotes the model parameter set. The luminosity distance is given by

\begin{equation}
d_L(z,\theta) = (1+z)c \int_{0}^{z} \frac{dz'}{H(z',\theta)},
\end{equation}
where $c$ represents the speed of light and $H(z,\theta)$ be the Hubble parameter determined by the chosen cosmological model. The Pantheon+SH0ES delivers the observed distance modulus $\mu_{\mathrm{obs}}(z_j)$ together with its covariance matrix $C$, which accounts for statistical as well as systematic uncertainties. The consistency between theoretical expectations and observations is quantified using the $\chi^2$ estimator

\begin{equation}
    \chi^2_{\mathrm{SN}} = \Delta \mu^{T} C^{-1} \Delta \mu,
\end{equation}
where, $\Delta \mu_j = \mu_{\mathrm{th}}(z_j,\theta) - \mu_{\mathrm{obs}}(z_j)$.

The {\bf Baryon Acoustic Oscillations} datasets arise from pressure-driven sound waves that propagated through tightly coupled photon-baryon plasma in the early Universe. These oscillations impart a characteristic scale in the large-scale distribution of the matter. This characteristic scale serves as a cosmological standard ruler for measuring distances and constraining the expansion history of the Universe. The physical scale associated with BAO is the sound horizon at the baryon drag epoch, $r_d$, which represents the maximum distance traveled by acoustic waves prior to photon-baryon decoupling. It is calculated as

\begin{equation}
r_d = \int_{z_d}^{\infty} \frac{c_s(z)}{H(z)} \, dz,
\end{equation}

where $c_s(z)$ is the sound speed of the photon–baryon plasma and $H(z)$ the Hubble expansion rate. Here, we use the BAO datasets DESI DR2 \cite{AbdulKarim_2025} survey. These datasets provide high-precision spectroscopic mapping of the large-scale structure of the Universe. In the first three years, DESI measured redshifts for tens of millions of galaxies and quasars, with a subset of high-quality measurements for BAO clustering analyses. The survey is performed in both bright and dark observational modes, targeting various galaxy populations over an extensive redshift range. This technique enables BAO measurements to be obtained over the approximate range $0.295 < z < 2.33$. It provides observational measurements of the scaled cosmological distance quantities $D_M/r_d$, $D_V/r_d$, and $D_H/r_d$, and can be expressed as

\begin{equation}
\frac{D_M}{r_d} = (1+z)^{-1}\frac{D_L}{r_d},
\qquad
\frac{D_V}{r_d} = \frac{\left(cD_L\frac{z}{H(z)} \right)^{\frac{1}{3}}}{r_d},
\qquad
\frac{D_H}{r_d} = \frac{c}{r_d H(z)},
\end{equation}

where $r_d$ denotes the sound horizon evaluated at the baryon drag epoch, $D_M$ represents the transverse comoving distance, $D_V$ corresponds to the volume-averaged distance and $D_H$ denotes the Hubble distance scale. We have performed the analysis using the DESI DR2 BAO dataset and a combined analysis of $H(z)$Pantheon+SH0ES and DESI DR2 BAO.

\begin{figure}[!htb] 
   \centering 
   \mbox{\includegraphics[scale=0.53]{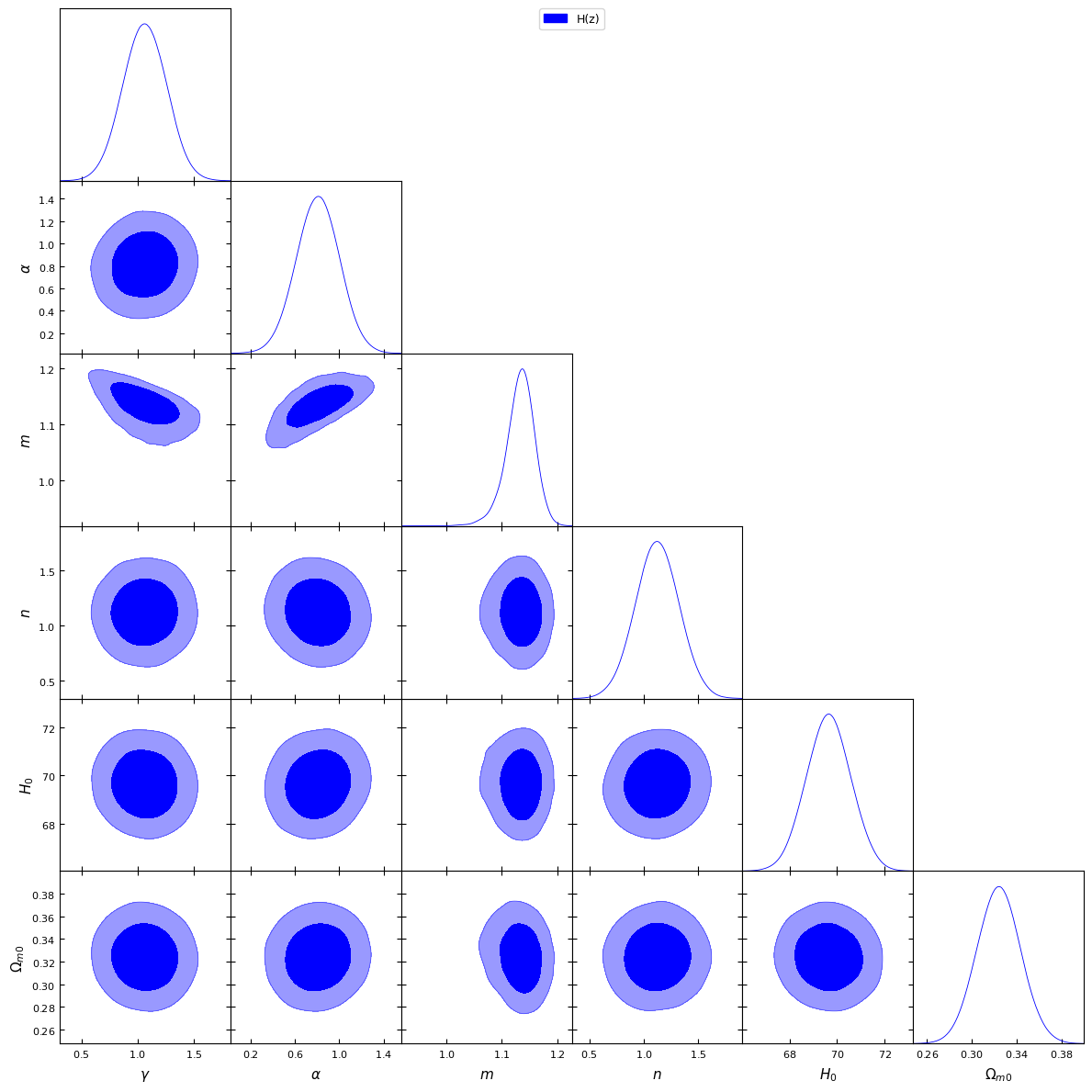}}   
\caption{\raggedright  A two-dimensional contour diagram derived from the observational $H(z)$ analysis, depicting best-fit parameter estimates and their confidence intervals for the model parameters $H_0$, $\gamma$, $\alpha$, $m$, $n$, and $\Omega_{m0}$ up to $2 \sigma$ level.} \label{MCMC with Hubble dataset}
\end{figure}

\begin{figure}[!htb] 
   \centering 
   \mbox{\includegraphics[scale=0.53]{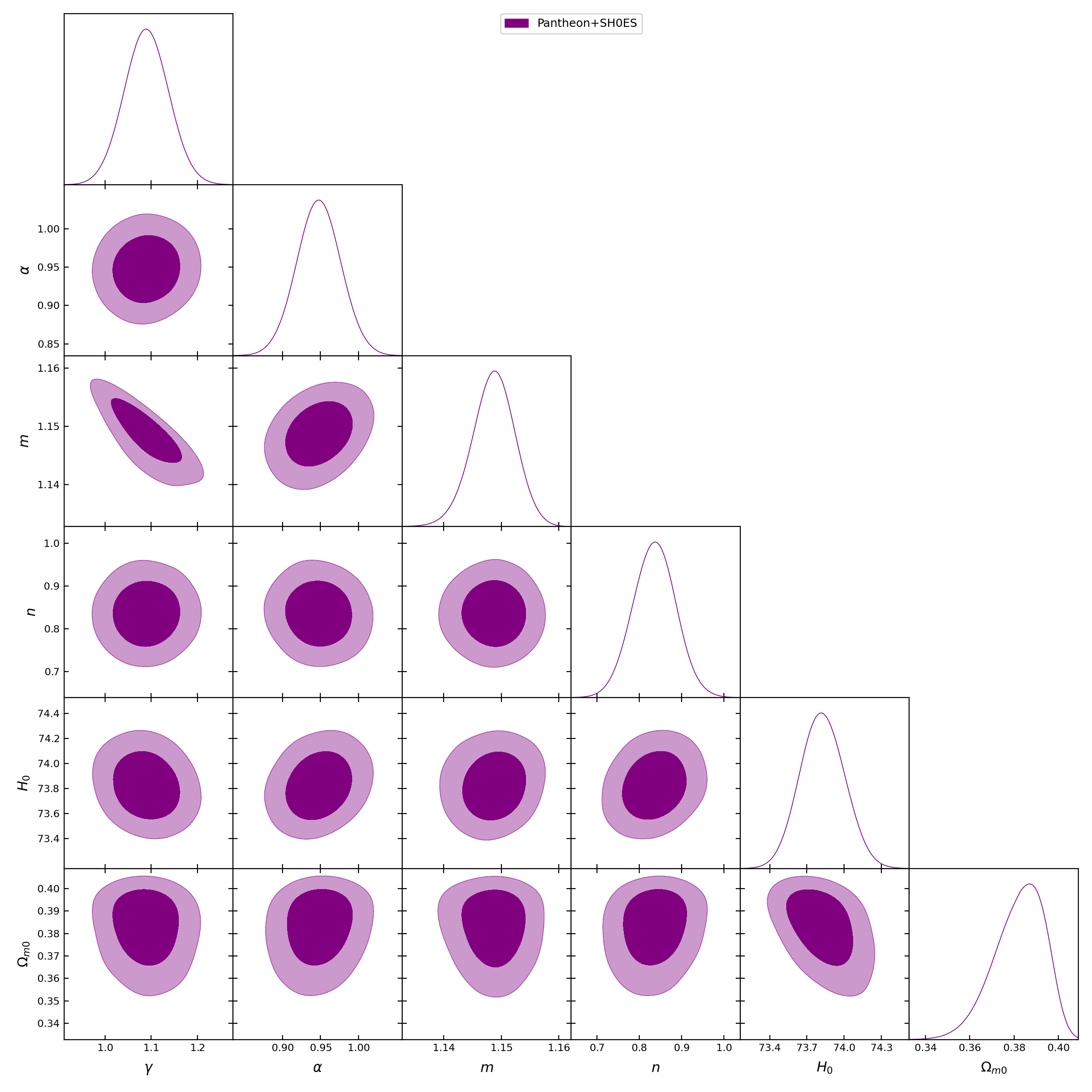}}   
\caption{\raggedright  A two-dimensional contour diagram derived from the Pantheon+SH0ES analysis, depicting best-fit parameter estimates and their confidence intervals for the model parameters $H_0$, $\gamma$, $\alpha$, $m$, $n$, and $\Omega_{m0}$ up to $2 \sigma$ level.} \label{MCMC with SN dataset}
\end{figure}

\begin{figure}[!htb] \label{DESI}
   \centering 
   \mbox{\includegraphics[scale=0.53]{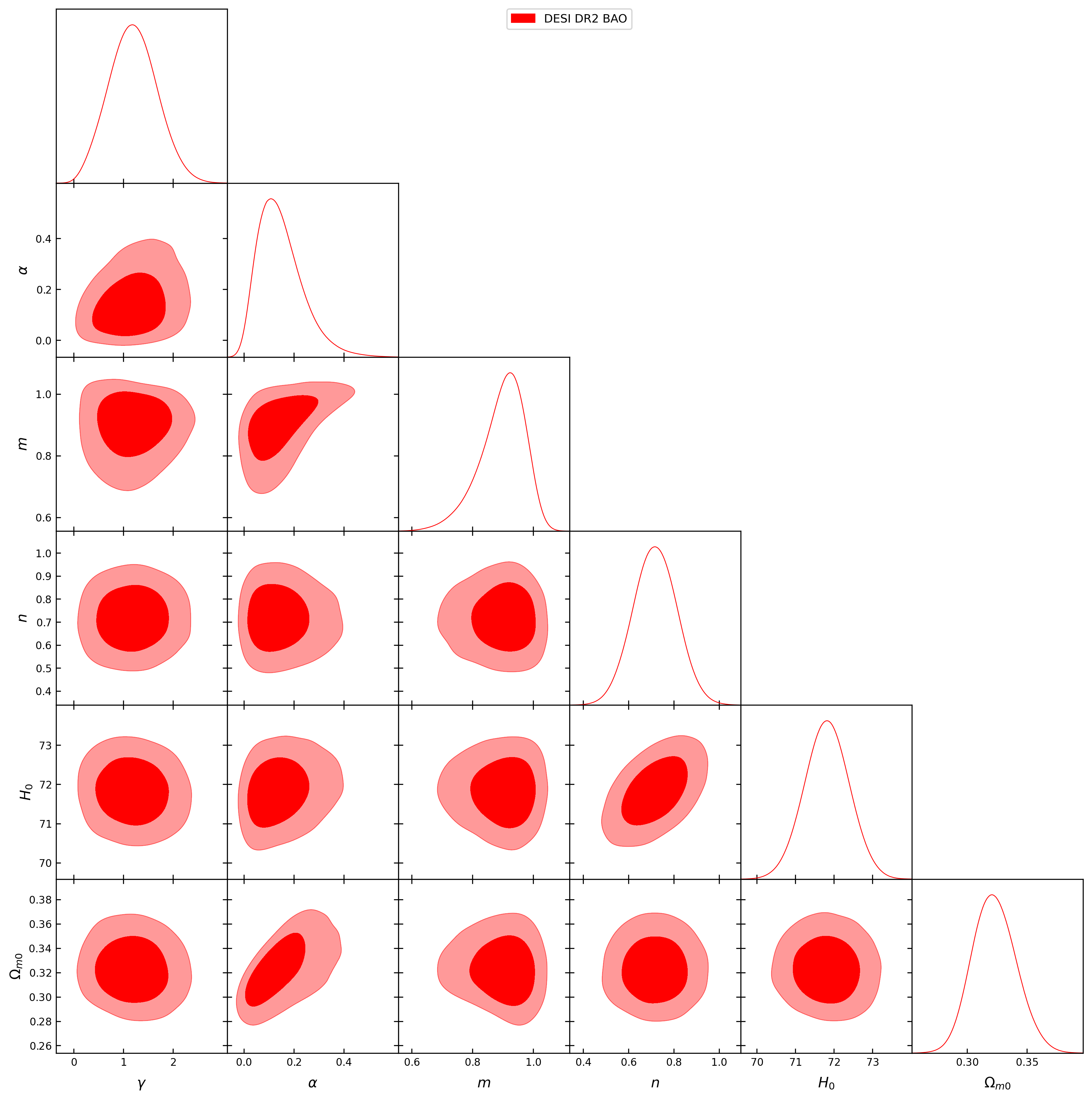}}   
\caption{\raggedright  A two-dimensional contour diagram derived from the DESI DR2 BAO analysis, depicting best-fit parameter estimates and their confidence intervals for the model parameters $H_0$, $\gamma$, $\alpha$ ,$m$, $n$, and $\Omega_{m0}$ up to $2 \sigma$ level.} \label{MCMC with DESI dataset}
\end{figure}

\begin{figure}[!htb]\label{combined}
   \centering 
   \mbox{\includegraphics[scale=0.53]{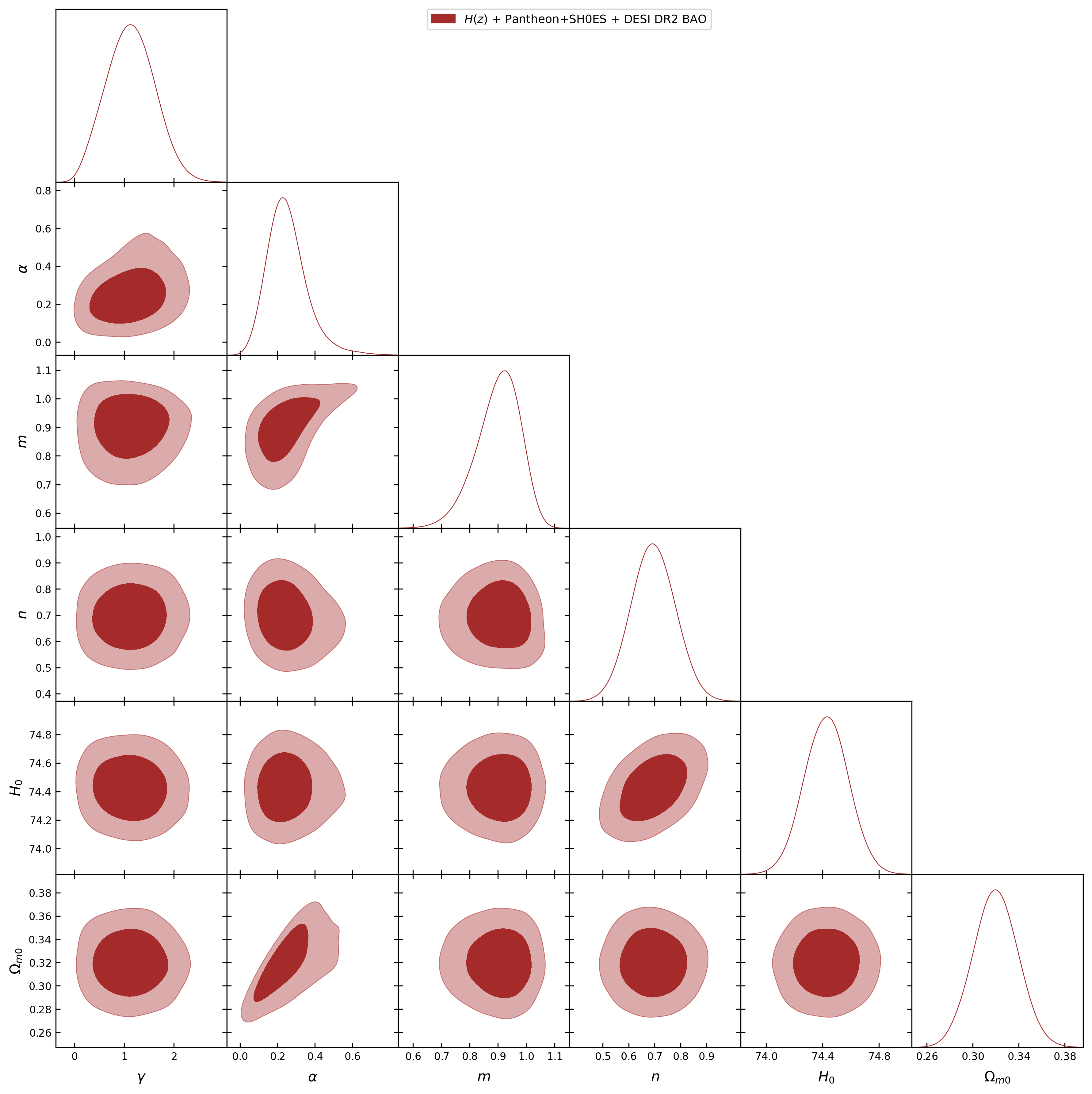}}   
    \caption{\raggedright  A two-dimensional contour diagram derived from the combined ($H(z)$ + Pantheon+SH0ES + DESI DR2 BAO) analysis, depicting best-fit parameter estimates and their confidence intervals for the model parameters $H_0$, $\gamma$, $\alpha$, $m$, $n$, and $\Omega_{m0}$ up to $2 \sigma$ level.}
   \label{MCMC with combined dataset}
\end{figure}

\begin{table*}
\centering
\caption{Confidence intervals for the model parameters obtained from different observational datasets.}
\label{constrained_Values_table}

\scriptsize
\setlength{\tabcolsep}{4pt}

\resizebox{\textwidth}{!}{%
\begin{tabular}{llcccccc}
\toprule
\textbf{Dataset} &  & 
$\boldsymbol{\gamma}$ &
$\boldsymbol{\alpha}$ &
$\boldsymbol{m}$ & 
$\boldsymbol{n}$ & 
$\boldsymbol{H_0}$ & 
$\boldsymbol{\Omega_{m0}}$ \\
\midrule

$H(z)$
 & $1\sigma$ & $1.06^{+0.19}_{-0.19}$ &
 $0.81^{+0.19}_{-0.20}$ &
 $1.13^{+0.02}_{-0.03}$ & $1.12^{+0.20}_{-0.20}$ & $69.64^{+0.93}_{-0.93}$ & $0.32^{+0.02}_{-0.02}$ \\
 & $2\sigma$ & $1.06^{+0.38}_{-0.38}$ &
 $0.81^{+0.38}_{-0.39}$ &
 $1.13^{+0.05}_{-0.06}$ & $1.12^{+0.40}_{-0.40}$ & $69.64^{+1.83}_{-1.81}$ & $0.32^{+0.04}_{-0.04}$ \\
\midrule

Pantheon+SH0ES
 & $1\sigma$ & $1.0897^{+0.0478}_{-0.0471}$ &
 $0.9477^{+0.0284}_{-0.0284}$ &
 $1.1487^{+0.0035}_{-0.0037}$ & $0.8362^{+0.0488}_{-0.0506}$ & $73.8217^{+0.1813}_{-0.1734}$ & $0.3840^{+0.0105}_{-0.0128}$ \\
 & $2\sigma$ & $1.0897^{+0.0944}_{-0.0930}$ &
 $0.9477^{+0.0570}_{-0.0569}$ &
 $1.1487^{+0.0070}_{-0.0077}$ & $0.8362^{+0.0985}_{-0.1005}$ & $73.8217^{+0.3513}_{-0.3356}$ & $0.3840^{+0.0151}_{-0.0263}$ \\
\midrule

DESI DR2 BAO
 & $1\sigma$ & $1.18^{+0.49}_{-0.48}$ &
 $0.13^{+0.10}_{-0.08}$ &
 $0.91^{+0.06}_{-0.08}$ & $0.72^{+0.10}_{-0.10}$ & $71.82^{+0.57}_{-0.58}$ & $0.32^{+0.02}_{-0.02}$ \\
 & $2\sigma$ & $1.18^{+1.00}_{-0.92}$ &
 $0.13^{+0.24}_{-0.12}$ &
 $0.91^{+0.11}_{-0.09}$ & $0.72^{+0.19}_{-0.19}$ & $71.82^{+1.16}_{-1.16}$ & $0.32^{+0.04}_{-0.03}$ \\
\midrule

$H(z)$ + Pantheon+SH0ES + DESI DR2 BAO
 & $1\sigma$ & $1.13^{+0.50}_{-0.49}$ &
 $0.24^{+0.11}_{-0.09}$ &
 $0.91^{+0.07}_{-0.09}$ & $0.69^{+0.09}_{-0.08}$ & $74.43^{+0.15}_{-0.16}$ & $0.32^{+0.02}_{-0.02}$ \\
 & $2\sigma$ & $1.13^{+0.97}_{-0.91}$ &
 $0.24^{+0.27}_{-0.17}$ &
 $0.91^{+0.12}_{-0.18}$ & $0.69^{+0.17}_{-0.16}$ & $74.43^{+0.31}_{-0.31}$ & $0.32^{+0.04}_{-0.04}$ \\
\bottomrule

\end{tabular}
}%
\end{table*}

\begin{figure}[!htb]
   \centering 
   \mbox{\includegraphics[scale=0.45]{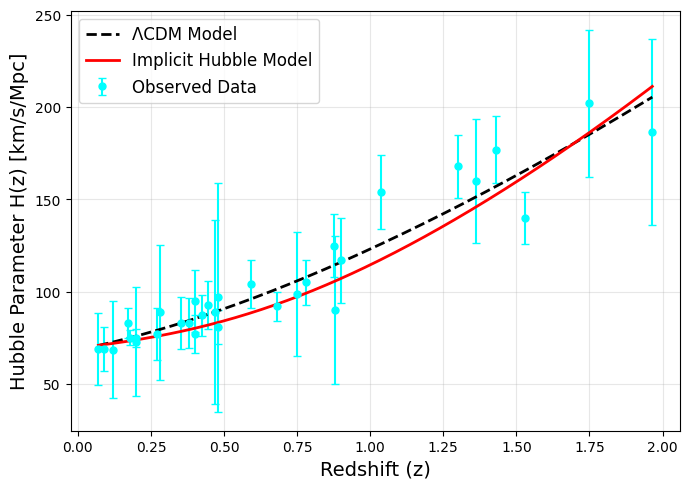}}   
    \hspace{10px}
    \mbox{\includegraphics[scale=0.45]{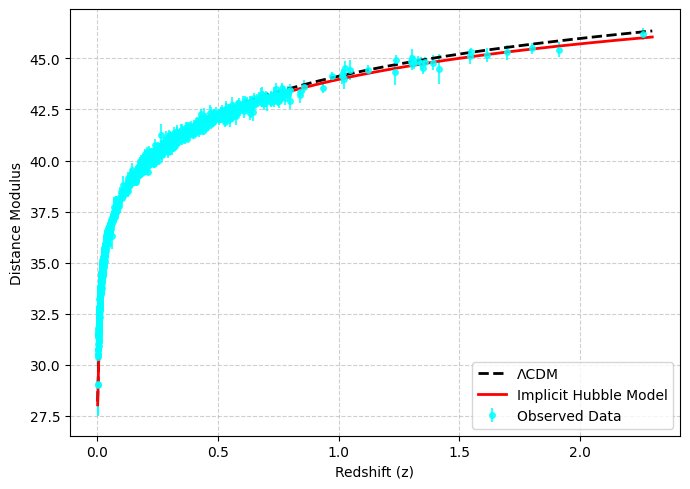}}
    \caption{\raggedright (i) \textbf{Left panel:} Redshift evolution of the Hubble parameter with corresponding error bars;
(ii) \textbf{Right panel:} Distance modulus as a function of redshift including observational uncertainties.}
   \label{errorbar}
\end{figure}

The contour plots for $H(z)$, Pantheon+SH0ES, DESI DR2 BAO and combination of all these are presented in Fig.--\ref{MCMC with Hubble dataset}, Fig.--\ref{MCMC with SN dataset},  
Fig.--\ref{MCMC with DESI dataset},  and Fig.--\ref{MCMC with combined dataset} respectively for $1\sigma$ and $2\sigma$ uncertainty regions. The observational constrained values of all the free parameters from the MCMC analysis that best fit the model are listed in Table--\ref{constrained_Values_table}. It can be noted that the range of $H_0$ and $\Omega_{m0}$ obtained is well within the range prescribed by different cosmological observations depicting the robustness of the model. While, the the values of $m$ and $n$ are slightly violating the conditions of Dynamical stability analysis for $H(z)$ and $n$ only for Pantheon+SH0ES data but satisfying for DESI DR2 BAO and Combination of all. So, the form of $f(T)$ and the corresponding model are able to reconstruct the accelerating behavior of the Universe. Further, Fig.--\ref{errorbar} displays the Hubble error bar plot and the distance modulus function extracted after fitting our implicit Hubble model with $H(z)$, the Pantheon+SH0ES dataset, respectively. It can be observed that the model curve is well-fitted with the error bars and in close alignment with the standard $\Lambda$CDM.

\section{Conclusion}\label{conclusion}
The cosmological dynamics of the well motivated functional form of $f(T)$ has been presented by performing the dynamical system analysis and using cosmological observations datasets. From the dynamical system analysis, we obtained a range of critical points that represents different cosmological eras such as radiation phase, matter phase and the accelerated expansion phase. The behavior of the phase space trajectories at the critical points shows that the trajectories at  $A_1$ are converging showing the stable behavior, while trajectories at $A_2$ and $A_3$ moving away confirm the unstable and saddle point behavior respectively. The evolutionary behavior shows that both the DE EoS and total EoS are converging to $\Lambda$CDM at late times. Further, the deceleration parameter shows transient behavior from early deceleration to late time acceleration. The present value of the matter density and dark energy density obtained is approximately $0.3$ and $0.7$ respectively. Our results show that the model allows for a viable transition from radiation to matter to an accelerated expansion phase of the Universe.\\

To validate the theoretical investigation, in the second phase, we performed the numerical simulations using the cosmological datasets and analyzed the late-time behavior of the Universe. The best-fit values of the parameters from $H(z)$, Pantheon+SH0ES, DESI DR2 BAO, and combined datasets are summarized in Table--\ref{constrained_Values_table}. The best-fit values for $H_0$ for $H(z)$, Pantheon+SH0ES, DESI DR2 BAO and combined datasets are respectively $69.64$, $73.8217$, $71.82$ and $74.43$; whereas the present value of  matter density parameter ($\Omega_{m0}$) are $0.32$, $0.3840$, $0.32$ and $0.32$. In addition, the range of the free parameters obtained as $m = 0.91^{+0.12}_{-0.18}$  $n = 0.69^{+0.17}_{-0.16}$ for combined datasets. It is interesting to note that the value of the free parameters obtained while finding the critical points and the range obtained from numerical simulation are in agreement. \\

Finally, we can conclude that the evolutionary behavior of the Universe is explored through dynamical stability analysis, and a stable accelerating cosmological model is achieved by integrating torsion instead of curvature and $f(T)$ form. The observational analysis strengthened this theoretical investigation by fitting the extracted Hubble model with various observational datasets.

\section*{Acknowledgement}  
BM acknowledges ANRF for the Mathematical Research Impact Centric Support (MATRICS)[File No: MTR/2023/000371].

\section{References}
\bibliographystyle{utphys}
\bibliography{biblio}
\end{document}